\documentclass[twocolumn,showpacs,amsmath,amssymb,10pt,aps,superscriptaddress,prl]{revtex4}
\usepackage{graphicx,color}
\usepackage{amsmath}
\usepackage{amssymb}
\usepackage{dsfont}
\usepackage{bbm}
\usepackage{color}

\usepackage[utf8]{inputenc}

\usepackage{amsfonts}
\usepackage{bm}

\newcommand{\ot}{{\,\otimes\,}}
\newcommand{{\Cd}}{{\mathbb{C}^d}}

\newcommand{\tr}{\mathrm{Tr}}

\newcommand{\BH}{\mathcal{B}(\mathcal{H})}

\def\<{\langle}
\def\>{\rangle}
\newtheorem{Theorem}{Theorem}
\newtheorem{Lemma}{Lemma}

\newtheorem{Corollary}{Corollary}

\newtheorem{Example}{Example}
\newtheorem{Proposition}{Proposition}

\newcommand{\beq}{\begin{equation}}
\newcommand{\eeq}{\end{equation}}
\newcommand{\bear}{\begin{eqnarray}}
\newcommand{\ear}{\end{eqnarray}}
\newcommand{\bdm}{\begin{displaymath}}
\newcommand{\edm}{\end{displaymath}}

\begin{document}

\title{Divisibility and Information Flow Notions of Quantum Markovianity for Noninvertible Dynamical Maps}
\author{Dariusz Chru\'sci\'nski}
\affiliation{Institute of Physics, Faculty of Physics, Astronomy and Informatics, Nicolaus Copernicus University, Grudzi\c{a}dzka 5/7, 87--100 Toru\'n, Poland}

\author{\'Angel Rivas}
\affiliation{Departamento de F\'isica Te\'orica I, Facultad de Ciencias F\'isicas, Universidad Complutense, 28040 Madrid, Spain.}

\affiliation{CCS-Center for Computational Simulation, Campus de Montegancedo UPM, 28660 Boadilla del Monte, Madrid, Spain}

\author{Erling St{\o}rmer}
\affiliation{Department of Mathematics, University of Oslo, 0316 Oslo, Norway}


\begin{abstract}

We analyze the relation between CP-divisibility and the lack of information backflow for an arbitrary -- not necessarily invertible -- dynamical map. It is well known that CP-divisibility always implies lack of information backflow. Moreover, these two notions are equivalent for invertible maps. In this letter it is shown that for a map which is not invertible the lack of information backflow always implies the existence of completely positive (CP) propagator which, however, needs not be trace-preserving. Interestingly, for a {\em wide class of image non-increasing dynamical maps} this propagator becomes trace-preserving as well and hence the lack of information backflow implies CP-divisibility. This result sheds new light into the structure of the time-local generators giving rise to CP-divisible evolutions. We show that if the map is not invertible then  positivity of dissipation/decoherence rates is no longer necessary for {CP-}divisibility.

\end{abstract}

\maketitle

{\em Introduction.}--- Recently, the notion of non-Markovian quantum evolution received considerable attention (see review papers \cite{NM1,NM2,NM3,WISE}). Quantum systems interacting with an environment \cite{Breuer,RivasHuelga} are of increasing relevance due to rapidly developing modern quantum  technologies like quantum communication or quantum computation {\cite{NC00}}. It turns out that recent experimental techniques allow us to go beyond standard Markovian approximations and observe new memory effects caused by the environmental interaction \cite{exp,exp2,exp3}. Therefore, there is a need for the exploration of non-Markovian regime, and characterizing the genuine properties of this kind of evolution.

To this end, two main approaches which turned out to be very influential are based on the concept of CP-divisibility \cite{RHP} and information flow \cite{BLP}. A dynamical map $\{\Lambda_t\}_{t\geq0}$ is a family of completely positive (CP) and trace-preserving (TP) maps acting on the space $\BH$ of bounded operators on the Hilbert space $\mathcal{H}$. In the present Letter we say that $\{\Lambda_t\}_{t\geq0}$ is \emph{divisible} if
\begin{equation}\label{div}
  \Lambda_t = V_{t,s} \Lambda_s,
\end{equation}
where $V_{t,s}:\BH \rightarrow \BH$ is a linear map for every $t \geq s$. Note that
since $\Lambda_t$ is TP the map $V_{t,s}$ is necessarily TP
 on the range of $\Lambda_s$ but needs not be
trace-preserving on the entire $\BH$. However, if $V_{t,s}$ is TP on $\BH$, one calls $\{\Lambda_t\}_{t\geq0}$ \emph{P-divisible} if $V_{t,s}$ is also a positive map on the entire $\BH$, and \emph{CP-divisible} if $V_{t,s}$ is CP on the entire $\BH$ \cite{footnote1}. According to \cite{RHP} the evolution is considered Markovian iff the corresponding dynamical map $\{\Lambda_t\}_{t\geq0}$ is CP-divisible. This definition is motivated by its classical limit, which is compatible with a classical Markovian process, and because such an evolution can be represented as the continuous limit of sequence of discrete interactions with a memoryless environment \cite{NM1,SM}.

A second idea is based on a physical feature of the system-reservoir interaction. It is claimed \cite{BLP} that the phenomenon of reservoir memory effects may be associated with an information backflow, that is, for any pair of density operators $\rho_1$ and $\rho_2$ one can define the information flow
\begin{equation}\label{BLP}
  \sigma(\rho_1,\rho_2;t) = \frac{d}{dt} \|\Lambda_t\rho_1 - \Lambda_t \rho_2\|_1 ,
\end{equation}
where $\|A\|_1$ denotes the trace norm of $A$. Following \cite{BLP} Markovian evolution is characterized by $\sigma(\rho_1,\rho_2;t) \leq 0$. Whenever $\sigma(\rho_1,\rho_2;t) > 0$ one calls it information backflow meaning that the information flows from the environment back to the system. In this case the evolution displays nontrivial memory effects and it is evidently non-Markovian.

Interestingly, both P- and CP-divisible maps have a clear mathematical characterization \cite{Angel}.

\begin{Theorem} \label{A} Let us assume that $\{\Lambda_t\}_{t\geq0}$ is an invertible dynamical map, i.e. $\Lambda_t^{-1}$ exists for any $t\geq0$. Then $\{\Lambda_t\}_{t\geq0}$ is P-divisible iff
\begin{equation}\label{P}
  \frac{d}{dt} \|\Lambda_t X\|_1 \leq  0 ,
\end{equation}
for any Hermitian $X \in \mathcal{B}(\mathcal{H})$. It is CP-divisible iff
\begin{equation}\label{CP}
  \frac{d}{dt} \|(\mathds{1} \ot \Lambda_t) {X}\|_1 \leq  0 ,
\end{equation}
for any Hermitian ${X} \in \mathcal{B}(\mathcal{H} \ot \mathcal{H})$.
\end{Theorem}
{Actually, CP- (or P-) divisibility implies (\ref{CP}) [or (\ref{P})] for an arbitrary map. Invertibility is only essential to prove the opposite implication.} Let us observe that the condition $\sigma(\rho_1,\rho_2;t) \leq 0$ is a slightly weaker version of (\ref{P}):  one takes $X = \rho_1-\rho_2$ which means that $X$ is Hermitian but traceless. In \cite{Angel} two of us proposed how to reconcile P-divisibility with information flow {by noticing that} any Hermitian operator $X$
can be interpreted (up to some multiplicative constant) as {a} so-called Helstrom matrix \cite{HEL} $X=p_1\rho_1-p_2\rho_2$ with $p_1+p_2=1$. It characterizes the error probability
of discriminating between states $\rho_1$ or $\rho_2$ with prior probabilities $p_1$ and $p_2$, respectively \cite{Angel}.

The relation between divisibility and information flow was recently reconsidered  by Bylicka {\it et al.} \cite{BOGNA}. They proved
\begin{Theorem} \label{B} Let $\{\Lambda_t\}_{t\geq0}$ be an invertible dynamical map, then it is CP-divisible iff
\begin{equation}\label{Bogna}
  \frac{d}{dt} \|(\mathds{1}_{d+1} \ot \Lambda_t)({\rho}_1-{\rho}_2)\|_1 \leq  0 ,
\end{equation}
for any pair of density operators ${\rho}_1$, ${\rho}_2$ in $\mathcal{B}(\mathcal{H}' \ot \mathcal{H})$ with $\dim(\mathcal{H}')-1=\dim(\mathcal{H})=d$.
\end{Theorem}
{Again, invertibility is only essential to prove that (\ref{Bogna}) implies CP-divisibility.}
So comparing (\ref{CP}) with (\ref{Bogna}) one enlarges the dimension of the ancilla $d \rightarrow d+1$,  but uses equal probabilities $p_1=p_2$, like in the original approach to the information flow \cite{BLP}.

If $t=0$ is the starting time for the system-environment interaction, any open system dynamics can be written as $\Lambda_t\rho={\rm Tr}_E[U(t,0)\rho\otimes\omega_E U^\dagger(t,0)]$ where $\omega_E$ is a fixed state of the environment. According to the postulates of  quantum mechanics $U(t,s)$ is a unitary evolution family which satisfies the Schr\"odinger Equation, and so it is continuous and differentiable. Since the partial trace is continuous but non-invertible, a dynamical map $\{\Lambda_t\}_{t\geq0}$ is a continuous, differentiable family (in the parameter $t$), but not necessarily invertible.

Interestingly,  Buscemi and {Datta} \cite{Datta}  analyzed  information backflow  defined in terms of the guessing probability of discriminating an ensemble of states $\{\rho_i\}$ $(i=1,2,\ldots $)  acting on $\mathcal{H}\ot \mathcal{H}$ with prior probabilities $p_i$. It was shown \cite{Datta} that a discrete time evolution is CP-divisible iff the guessing probability decreases for any ensemble of states. In this approach invertibility of the maps plays no role and hence this approach is universal. However, the price one pays, is the use of ensembles containing arbitrary number of states $\rho_i$, which makes the whole approach hardly implementable. Moreover, since just a discrete evolution $\Lambda_n$ is considered, there is not direct relation to the problem for continuous dynamical maps. For example such maps do not satisfy time-local master equations. Anyway, \cite{Datta}  poses an important question {\em whether the assumption of invertibility in Theorem \ref{A} may be removed}.

{In this Letter we show how to generalize Theorem \ref{A} and \ref{B} to non-invertible dynamical maps.} This result sheds  new light into time-local master equations
\begin{equation}\label{}
  \frac{d}{dt} \Lambda_t = \mathcal{L}_t \Lambda_t \ , \ \Lambda_{t=0} = \mathds{1} .
\end{equation}
One usually says that the corresponding solution $\{\Lambda_t\}_{t\geq0}$ is CP-divisible if the two-point propagator
\begin{equation}\label{}
  V_{t,s} = \mathcal{T} e^{\int_s^t \mathcal{L}_\tau d\tau } ,
\end{equation}
is CPTP for any $t \geq s$, and hence one concludes that $\mathcal{L}_t$ is a time-dependent GKLS generator \cite{GKLS}.  However, it turns out to be true only for invertible dynamics. In this Letter we show that if $\{\Lambda_t\}_{t\geq0}$ is not invertible, it can still be CP-divisible even if the corresponding generator $\mathcal{L}_t$ does not have GKLS structure.

\vspace{.2cm}

{\em Divisible maps.}--- Interestingly, the property of divisibility is fully characterized by the following

\begin{Proposition} \label{propDiv1} A dynamical map $\{\Lambda_t\}_{t\geq0}$ is divisible iff
\begin{equation}\label{KER}
  {\rm Ker}(\Lambda_t) \supseteq  {\rm Ker}(\Lambda_s),
\end{equation}
for any $t > s$.
\end{Proposition}
Proof: {If $\{\Lambda_t\}_{t\geq0}$ is divisible and $X \in {\rm Ker}(\Lambda_s)$, then
\begin{equation*}\label{}
   \Lambda_tX = V_{t,s} (\Lambda_s X) = V_{t,s}0 = 0,
\end{equation*}
and hence $X \in {\rm Ker}(\Lambda_t)$.

Suppose now that (\ref{KER}) is satisfied. To show that $\{\Lambda_t\}_{t\geq0}$ is divisible we provide a construction for $V_{t,s}$. This construction is highly non-unique: if $Y \in {\rm Im}(\Lambda_s)$, i.e. there exists $X$ such $\Lambda_s X = Y$, we define $V_{t,s} Y = \Lambda_tX$. Suppose now that $Y \notin {\rm Im}(\Lambda_s)$ and let $\Pi_s : \BH \rightarrow {\rm Im}(\Lambda_s)$ be a (Hermiticity preserving) projector onto ${\rm Im}(\Lambda_s)$ \cite{SM}, that is, $\Pi_s \Pi_s = \Pi_s$ is an identity operation on ${\rm Im}(\Lambda_s)$. Define
\begin{equation}\label{}
  V_{t,s} Y := \Lambda_t X ,
\end{equation}
where $X$ is an arbitrary element such that $\Pi_s Y = \Lambda_s X$. It only remains to prove that this is a well-defined construction. Indeed, if $\Lambda_s X = \Lambda_s X' = \Pi_sY$, then our construction implies $\Lambda_t X = \Lambda_t X'$ for $t > s$. Specifically, $\Delta = X - X' \in {\rm Ker}(\Lambda_s)$ and hence due to (\ref{KER}) one has $  \Delta  \in {\rm Ker}(\Lambda_t)$ which implies $\Lambda_t \Delta = \Lambda_t X - \Lambda_t X' =0$. It should be stressed, however, that $V_{t,s}$ needs not be TP due to the fact that the projector $\Pi_s$ needs not be TP.

\par \hfill $\Box$ }

Note that if $\{\Lambda_t\}_{t\geq0}$ is invertible, then it is always divisible due to $V_{t,s} = \Lambda_t \Lambda_s^{-1}$. In this case condition (\ref{KER}) is trivially satisfied: ${\rm Ker}(\Lambda_t) =  {\rm Ker}(\Lambda_s) =0$.

Actually, there is a simple sufficient condition for divisibility

\begin{Proposition}\label{propDiv2} If the dynamical map $\{\Lambda_t\}_{t\geq0}$ satisfies
\begin{equation}\label{P1}
  \frac{d}{dt} \| \Lambda_t X\|_1 \leq 0 ,
\end{equation}
for all Hermitian $X \in \BH$, then it is divisible.
\end{Proposition}
Proof: Suppose that \eqref{P1} is satisfied but $\{\Lambda_t\}_{t\geq0}$ is not divisible, that is, by Proposition \ref{propDiv1} there exists $X$ such that $\Lambda_s X=0$ but $\Lambda_t X\neq 0$ $(t>s)$. This shows $\|\Lambda_t X\|_1 > 0=\|\Lambda_s X\|_1$ and hence $ \| \Lambda_t X\|_1$ does not monotonically decrease.
\par \hfill $\Box$

Clearly, the above condition is sufficient but not necessary, since any invertible $\{\Lambda_t\}_{t\geq0}$ is divisible even if it does not satisfy (\ref{P1}).

\vspace{.2cm}

{\em Arbitrary dynamical maps.}--- Now we prove the central result which provides generalizations of Theorems \ref{A} and \ref{B} for arbitrary, that is, not necessarily invertible, dynamical maps. Let us start with a pair of lemmas.

\begin{Lemma} \label{L}
Let $M$ be a linear subspace in $\mathcal{B}(\mathcal{H})$, and consider a trace-preserving linear map $\Phi : M \to \mathcal{B}(\mathcal{H})$. If $\Phi$ is a contraction in the trace norm, then it is positive.
\end{Lemma}
Proof: take arbitrary $X\geq 0$ from $M$. One has $\|X\|_1 = {\rm Tr}(X)$. Now, since $\Phi$ is trace-preserving ${\rm Tr}(X) = {\rm Tr}[\Phi(X)] \leq {\rm Tr}|\Phi(X)| = \|\Phi(X)\|_1$. Finally, since $\Phi$ is a contraction $\|\Phi(X)\|_1 \leq \|X\|_1$, so that
\begin{equation}\label{}
  \|\Phi(X)\|_1 = {\rm Tr}[\Phi(X)] ,
\end{equation}
which proves that $\Phi(X) \geq 0$. \hfill $\Box$\\

Note, that if $M = \mathcal{B}(\mathcal{H})$ then one recovers the well known result \cite{Kossak,Ruskai} used in \cite{Angel} and recently in \cite{BOGNA}.

\begin{Lemma} \label{L2}
Let $M$ be a linear subspace in $\mathcal{B}(\mathcal{H})$ with $\dim(\mathcal{H})=d$. If $M$ is spanned by positive operators (density matrices), then a $d$-positive map $\Phi : M \to \mathcal{B}(\mathcal{H})$ can be extended to a CP map $\widetilde{\Phi} :  \mathcal{B}(\mathcal{H}) \to \mathcal{B}(\mathcal{H})$.
\end{Lemma}

The problem of CP extensions of a CP map $\Phi : M \to \mathcal{B}(\mathcal{H})$ is well-studied in the theory of operator algebras and was solved by Arveson \cite{Arveson} when $M$ defines an operator system (see also \cite{Paulsen,Erling}). Recently the extension problem was studied in the context of quantum operations in \cite{Jencova,Teiko} beyond operator systems. In particular, Jencova proves Lemma \ref{L2} in \cite{Jencova}. Nevertheless, for the sake of completennes we include a explicit proof in the supplementary material \cite{SM}. 

\begin{Theorem} \label{C} {If a dynamical map} $\{\Lambda_t\}_{t\geq0}$ satisfies
\begin{equation}\label{dCPdiv}
  \frac{d}{dt} \|(\mathds{1} \ot \Lambda_t) {X}\|_1 \leq  0 ,
\end{equation}
for any Hermitian ${X} \in \mathcal{B}(\mathcal{H} \ot \mathcal{H})$, then it is divisible with CP propagators $V_{t,s}$.
\end{Theorem}
Proof: By Proposition 2 the dynamical map $\{\mathds{1} \otimes \Lambda_t\}_{t\geq0}$ is divisible, hence so is $\{\Lambda_t\}_{t\geq0}$, therefore $\Lambda_t = V_{t,s}\Lambda_s$. If the map $\Lambda_s$ is not invertible, the propagator $V_{t,s}$ is not uniquely defined. We show that one can find $V_{t,s}$ which is CP. Note, that (\ref{dCPdiv}) implies that $\mathds{1} \ot V_{t,s}$ is a contraction on the image of $\mathds{1} \ot \Lambda_s$ \cite{Angel,BOGNA}. Since $\mathds{1} \ot V_{t,s}$ is trace-preserving on  ${\rm Im}(\mathds{1} \ot \Lambda_s)$
Lemma \ref{L} implies that $\mathds{1} \ot V_{t,s}$ is positive on  ${\rm Im}(\mathds{1} \ot \Lambda_s)$ or equivalently that $V_{t,s}$ is $d$-positive on  ${\rm Im}(\Lambda_s)$. It should be stressed, that $V_{t,s}$ is defined on the linear subspace ${\rm Im}(\Lambda_s) \subset \mathcal{B}(\mathcal{H})$. Now, the question is about the extension of $V_{t,s}$ to the whole operator space $\mathcal{B}(\mathcal{H})$. However, since the subspace ${\rm Im}(\Lambda_s)$ is spanned by the positive operators $\Lambda_s(X)$, where $X$ are positive operators from $\mathcal{B}(\mathcal{H})$, Lemma \ref{L2} guaranties the existence of a CP extension
$\widetilde{V}_{t,s} : \mathcal{B}(\mathcal{H}) \to \mathcal{B}(\mathcal{H})$. One has, therefore,
\begin{equation}\label{}
  \Lambda_t  = V_{t,s} \Lambda_s = \widetilde{V}_{t,s}  \Lambda_{s} .
\end{equation}
\hfill $\Box$

Clearly, if $\Lambda_s$ is invertible then $\widetilde{V}_{t,s}  = V_{t,s}$. It should be stressed, however, that generically \emph{$\widetilde{V}_{t,s}$ needs not be trace-preserving}. It is always trace-preserving on ${\rm Im}(\Lambda_s)$. Hence, monotonicity property (\ref{dCPdiv}) does not imply CP-divisibility but a slightly weaker property.  Examples of CP extensions which are not trace-preserving were recently provided in \cite{Teiko}.

\vspace{.2cm}

{\em Image non-increasing dynamical maps.}--- Consider now a wide class of dynamical maps which satisfy
\begin{equation}\label{ker-im}
  {\rm Im}(\Lambda_t) \subseteq  {\rm Im}(\Lambda_s), \quad t> s.
\end{equation}
We shall refer to these as ``image non-increasing dynamical maps''. Note that ``kernel non-decreasing" Eq. \eqref{KER} (equivalent to divisibility) only implies ${\rm dim} [{\rm Im}(\Lambda_t)] \leq  {\rm dim}[{\rm Im}(\Lambda_s)]$. 
There are two natural classes of maps satisfying (\ref{ker-im}). The first class are normal divisible maps, i.~e.
 $ \Lambda_t \Lambda_t^\dagger =  \Lambda_t^\dagger\Lambda_t$, 
where $\Lambda_t^\dagger$ is the dual map (Heisenberg picture), that is, ${\rm Tr}[\Lambda_t^\dagger(X)\rho] = {\rm Tr}[X\Lambda_t(\rho)]$.  For normal maps the kernel is orthogonal to the image, so divisibility implies \eqref{KER} and hence (\ref{ker-im}) immediately follows. The second class are diagonalizable commutative maps (here commutative means $\Lambda_t \Lambda_s = \Lambda_s \Lambda_t$ for arbitrary $t$ and $s$).   In this case $\Lambda_t$ is characterized by the diagonal representation
 $ \Lambda_t\rho = \sum_\alpha \lambda_\alpha(t) F_\alpha {\rm Tr}(G_\alpha^\dagger \rho)$,
with time independent damping basis  \cite{DB} $\{F_\alpha,G_\beta\}$ such that ${\rm Tr}(F^\dagger_\alpha G_\beta) = \delta_{\alpha\beta}$ $(\alpha,\beta=0,1,\ldots,d^2-1)$. 

\begin{Theorem} \label{D} If  the image non-increasing dynamical map $\{\Lambda_t\}_{t\geq0}$  satisfies
\begin{equation}\label{CP-diag}
  \frac{d}{dt} \|(\mathds{1} \ot \Lambda_t) X\|_1 \leq  0 ,
\end{equation}
for any Hermitian ${X} \in \mathcal{B}(\mathcal{H} \ot \mathcal{H})$, then it is  CP-divisible.
\end{Theorem}
Proof: clearly (\ref{CP}) implies (Theorem \ref{C}) that $\{\Lambda_t\}_{t\geq0}$ is divisible with $V_{t,s}$ which is CPTP on ${\rm Im}(\Lambda_s)$. Since $\Lambda_{t=0}=\mathds{1}$, continuity implies that there exists some small $\epsilon$ such that $\Lambda_{\epsilon}$ is invertible. Let us take $t_1$ the smallest time instant where the dynamics becomes non-invertible, i.~e. $\{\Lambda_t\}_{t_1>t\geq0}$ is invertible. Then we can write
$\Lambda_{t_1}=V_{t_1,t_1-\epsilon}\Lambda_{t_1-\epsilon}$, 
where $V_{t_1,t_1-\epsilon}$ is CPTP [on the entire $\BH$] for $\epsilon\in(0,t_1)$.
Consider now the operator
\begin{equation}\label{proj1}
\Pi_{t_1}:=\lim_{\epsilon\rightarrow0^+}V_{t_1,t_1-\epsilon}.
\end{equation}
It turns out that $\Pi_{t_1}$ is a CPTP projection onto ${\rm Im}(\Lambda_{t_1})$. We provide a detailed proof of this in the supplementary material \cite{SM}. Hence $\widetilde{V}_{t,t_1} = V_{t,t_1} \Pi_{t_1}$ is CPTP on the entire $\BH$. Consider now the smallest $t_2 > t_1$ such that ${\rm dim} [{\rm Im}(\Lambda_{t_2})] <   {\rm dim} [{\rm Im}(\Lambda_{t_1})]$ and  ${\rm dim} [{\rm Im}(\Lambda_t)] = {\rm dim} [{\rm Im}(\Lambda_{t_1})]$ for $t_1\leq t < t_2$. For image non-increasing dynamical maps it means that
${\rm Im}(\Lambda_{t_2}) \subset    {\rm Im}(\Lambda_{t_1})$, and ${\rm Im}(\Lambda_t) =   {\rm Im}(\Lambda_{t_1})$ for $t_1\leq t < {t_2}$.  Then considering $\{V_{t,s}\}_{t_2>t>s\geq t_1}$ as a bijective family of maps on the space ${\rm Im}(\Lambda_{t_1})$, the same argument as before, with the role of $\Lambda_t$ played now by $V_{t,t_1}$, applies to show that $\Pi_{t_2}$ is a CPTP on ${\rm Im}(\Lambda_{t_1})$, which projects onto ${\rm Im}(\Lambda_{t_2})$. Finally, let $\{t_1,\ldots,t_k\}$ be a set such that ${\rm dim}[{\rm Im}(\Lambda_t)]$ is discontinuous, that is,
\begin{equation*}
{\rm dim}[{\rm Im}(\Lambda_{t_1})] >    {\rm dim}[{\rm Im}(\Lambda_{t_2})] > \ldots > {\rm dim}[{\rm Im}(\Lambda_{t_k})].
\end{equation*}
Note, that for $t \in [t_i,t_{i+1})$ one has
\begin{equation}\label{}
{\rm Im}(\Lambda_{t_{i}})= {\rm Im}(\Lambda_{t})\supsetneq {\rm Im}(\Lambda_{t_{i+1}}).
\end{equation}
Hence, for $s \in [t_i,t_{i+1})$ one defines
\begin{equation}\label{}
  \widetilde{V}_{t,s} = V_{t,s}  \Pi_{t_{i}} \ldots \Pi_{t_{1}} ,
\end{equation}
which is CPTP on the entire $\BH$. \hfill $\Box$

Note that a parallel argument applies to show the equivalence between Eq. \eqref{P1} and P-divisibility in the case of image non-increasing maps.

\begin{Theorem}: \label{bognaup} A dynamical map $\{\Lambda_t\}_{t\geq0}$  satisfying
\begin{equation}\label{dCPdiv2}
\frac{d}{dt}\|(\mathds{1}_{d+1} \otimes\Lambda_t)({\rho}_1-{\rho}_2)\|_1\leq0
\end{equation}
for any pair of density operators ${\rho}_1$, ${\rho}_2$ in $\mathcal{B}(\mathcal{H}' \ot \mathcal{H})$ with $\dim(\mathcal{H}')-1=\dim(\mathcal{H})=d$ is divisible with $CP$ propagators $V_{t,s}$. In addition, if the map is image non-increasing, it is CP-divisible.
\end{Theorem}
The proof of this theorem follows from Theorems \ref{C} and \ref{D}, and a similar argument as in \cite{BOGNA}. We leave it as a part of the supplementary material \cite{SM}.

\vspace{.2cm}

{\em CP-divisibility vs. master equation} -- Any differentiable $\Lambda_t$ satisfies a time-local master equation of the form of
\begin{equation}\label{ME}
  \frac{d}{dt} {\Lambda}_t = \mathcal{L}_t \Lambda_t, \ \ \Lambda_0 = \mathds{1} ,
\end{equation}
so that $V_{t,s} = \mathcal{T} e^{\int_s^t \mathcal{L}_\tau d\tau }$. Then CP-divisibility implies that $V_{s,s}$ is a CPTP identity map on some subspace $M$, such that ${\rm Im}(\Lambda_s)\subseteq M \subseteq \BH$. Moreover, if $\{\mathds{1}\otimes\Lambda_t\}_{t\geq0}$ is image non-increasing and contracting, there exists a CPTP projector $\Pi_s$ onto ${\rm Im}(\Lambda_s)$.

\begin{Corollary} If the image non-increasing dynamical map $\{\Lambda_t\}_{t\geq0}$ satisfies a time-local master equation (\ref{ME}), then it is CP-divisible iff $\mathds{1} \ot \mathcal{T} e^{\int_s^t \mathcal{L}_\tau d\tau } $ is a TP contraction on $\BH \ot {\rm Im}(\Lambda_s)$ for all pairs $t \geq s$.
\end{Corollary}

In the following examples we will show that this does not require a time dependent GKLS form for all times (another example can be found in \cite{SM}).

\begin{Example}[Amplitude damping channel]
The dynamics of a single amplitude-damped qubit is
governed by a single function $G(t)$ which depends on the form of the reservoir spectral density $J(\omega)$ \cite{Breuer}:{
\begin{equation}\label{}
  \Lambda_t \rho = \left( \begin{array}{cc} |G(t)|^2\rho_{11} & G(t) \rho_{12} \\ G^*(t) \rho_{21} & (1-|G(t)|^2)\rho_{11}+ \rho_{22} \end{array} \right) ,
\end{equation}}
This evolution is generated by the following time-local generator
\begin{equation*}\label{}
  \mathcal{L}_t \rho = - \frac{is(t)}{2}[\sigma_+\sigma_-,\rho] + \gamma(t)( \sigma_- \rho\sigma_+ - \frac 12 \{ \sigma_+\sigma_-,\rho\} ) ,
\end{equation*}
where $\sigma_\pm$ are the spin lowering and rising operators together with
$s(t) = -2{\rm Im} \frac{\dot{G}(t)}{G(t)}$, and $\gamma(t) = -2{\rm Re} \frac{\dot{G}(t)}{G(t)}$.
{This generator is commutative and diagonalizable}. Now, the dynamical map is invertible whenever $G(t) \neq 0$. Suppose now that $G(t_*)=0$ and $G(t) \neq 0$ for $t < t_*$ (note that $G(0)=1$). The image of $\Lambda_{t_*}$ is just proportional to the ground state $P_0 = \sigma_- \sigma_+$, and so a CPTP projector onto ${\rm Im}(\Lambda_{t_*})$  reads
\begin{equation}\label{}
  \Pi_{t_*} X = P_0 {\rm Tr} X .
\end{equation}
It is, therefore clear that $\{\Lambda_t\}_{t\geq0}$ is divisible iff $G(t)=0$ for $t \geq t_*$. Hence, finally, the map $\{\Lambda_t\}_{t\geq0}$ is CP-divisible iff it is divisible and $\gamma(t) \geq 0$ for $t < t_*$. Note that $\gamma(t)$ is arbitrary for $t\geq t_*$. The only constraint is $G(t) = 0$:  $\gamma(t)$ blows up to $+\infty$ at $t=t_*$, and then is arbitrary provided $\int_0^{t}\gamma(\tau)d\tau = \infty$ for all $t \geq t_*$.
Hence, positivity of $\gamma(t)$ is sufficient but not necessary for CP-divisibility. It is necessary only if $\gamma(t)$ is finite for finite times, that is, the generator $\mathcal{L}_t$ is regular and the map $\Lambda_t$ is invertible. Note that divisibility means that if the system relaxed to the ground state (at time $t_*$) it stays in that state forever. In addition, CP-divisibility means that the relaxation to the ground state was monotonic $\frac{d}{dt}|G(t)|\leq 0$.
\end{Example}

\begin{Example}[Random unitary evolution] \label{E1} Consider the qubit evolution governed by the following time-local generator
\begin{equation}\label{Pauli}
  \mathcal{L}_t \rho = \frac 12 \sum_{k=1}^3  \gamma_k(t)( \sigma_k \rho \sigma_k - \rho) ,
\end{equation}
which leads to the unital dynamical map (time-dependent Pauli channel): $\Lambda_t\rho =  \sum_{\alpha=0}^3  p_\alpha(t) \sigma_\alpha \rho \sigma_\alpha $. The map is invertible if its corresponding eigenvalues
\[
\lambda_i(t) = e^{-\Gamma_j(t) - \Gamma_k(t)} \ ;\ \  \Gamma_j=\int_0^t \gamma_j(\tau)d\tau
\]
are different from zero. Here $\{i,j,l\}$ is a permutation of $\{1,2,3\}$ (note, that the remaining eigenvalue $\lambda_0(t)=1$).
Now, if for example $\Gamma_3(t_1) = \infty$ at finite time $t_1$, then $\lambda_1(t_1)=\lambda_2(t_1)=0$, and hence divisibility implies  $\lambda_1(t)=\lambda_2(t)=0$ for $t\geq t_1$. One finds the corresponding CPTP projector
\begin{equation*}
  \Pi_{t_1} X =  \frac 12 ( X +  \sigma_3 X \sigma_3 ) .
\end{equation*}
Note that $ \Pi_{t_1} \sigma_1 = \Pi_{t_1} \sigma_2 =0$. Now, if at $t_2 > t_1$ one has $\Gamma_2(t_2)=\infty$ (or equivalently  $\Gamma_1(t_2)=\infty$), then $\lambda_3(t_2)$ vanishes as well and hence divisibility implies $\lambda_i(t)=0$ for $t \geq t_2$ ($i=1,2,3$).
One finds the corresponding CPTP projector
\begin{equation*}
  \Pi_{t_2} X =  \frac 12 \mathbb{I}\, {\rm Tr}(X) ,
\end{equation*}
that is, it fully depolarizes an arbitrary state of the system. To summarize: the evolution is CP-divisible iff all $\gamma_\alpha(t) \geq 0$ for $t < t_1$, and $\gamma_3(t)$ continues to be nonegative up to $t_2$. From $t_2$ on the system stays at the maximally mixed state.
\end{Example}

\vspace{.2cm}

{\em Conclusions.} --- In this Letter we analyzed the relation between monotonicity of the trace norm (\ref{dCPdiv}) and CP-divisibility of the dynamical map $\{\Lambda_t\}_{t\geq0}$. While CP-divisibility always implies (\ref{dCPdiv}), it is well known that for invertible maps the converse is also true, that is, these two notions are equivalent. For maps which are not invertible the situation is much more subtle (as was recently noticed in \cite{BOGNA}). We proved that in this case, Eq. (\ref{dCPdiv}) implies a slightly weaker property --- there exists a family of completely  positive maps $V_{t,s}$ on $\BH$ which are trace-preserving on the image of $\Lambda_s$ [but not on the entire $\BH$]. Interestingly, for maps which are image non-increasing trace-preservation is guarantied on $\BH$ and hence they are CP-divisible.  Notably this result sheds new light into the structure of the time-local generator $\mathcal{L}_t$ which gives rise to CP-divisible evolution. For invertible maps, $\mathcal{L}_t$ has a structure of time-dependent GKLS generator, in particular all dissipation rates  $\gamma_k(t) \geq 0$ for all $t \geq 0$ \cite{Canonical}. It is no longer true for dynamical maps which are not invertible, that is, they  correspond to singular generators {\cite{Hall}}. In this case $\gamma_k(t) \geq 0$ but only for $t \in [0,t_*)$, where $t_*$ is the first moment of time where $\Lambda_t$ becomes non-invertible. For $t \geq t_*$ some $\gamma_k(t)$ might be temporarily negative, and still the evolution might be CP-divisible. The point $t_*$ at which some $\gamma_k(t)$ becomes singular, provides an analog of the event horizon: the future behavior of a set of $\gamma_k(t)$ does not effect the evolution of the system. A typical example is the evolution reaching equilibrium state at finite time $t_*$. Then the system stays at equilibrium forever irrespective of the future ($t > t_*$) time dependence of the generator.

Finally, we note that the relation between the result by Buscemi and Datta \cite{Datta} on guessing probabilities for discrete evolution $\Lambda_n$ and our results on continuous evolution is not evident and deserves further analysis. On the other hand, the general problem of finding a CPTP extension of a CPTP propagator $V_{t,s}$ on a subspace remains open. If possible, it would ensure the complete equivalence of CP-divisibility and complete contractivity.

\begin{acknowledgments}

DC was supported by the National Science Center project 2015/17/B/ST2/02026. AR acknowledges the Spanish MINECO grants FIS2015-67411, FIS2012-33152, the CAM research consortium QUITEMAD+ S2013/ICE-2801, and U.S. Army Research Office through grant W911NF-14-1-0103 for partial financial support.
We thank Marcus Johansson for interesting discussion and pointing out an error in the first version of the manuscript. Many thanks to Adam Skalski and Vern Paulsen for valuable comments.

\end{acknowledgments}

\onecolumngrid
\clearpage
\twocolumngrid

\section*{SUPLEMENTARY MATERIAL}

\appendix

\setcounter{figure}{0}
\setcounter{equation}{0}
\renewcommand*{\thefigure}{S\arabic{figure}}
\renewcommand*{\theequation}{S\arabic{equation}}

\subsection{I.\quad Connection between Markovianity and P/CP-divisibility for non-invertible maps}
\label{app_A0}
The Markovianity definition in \cite{RHP_SM} requires $V_{t,s}$ to be a CPTP map on $\mathcal{B}(\mathcal{H})$, and not only on ${\rm Im}(\Lambda_s)$. There are, at least, two reasons to motivate this.
\smallskip
\begin{itemize}
\item \underline{Correct Classical Limit}:
Markov processes are unambiguously defined for classical random variables. They are the ones satisfying the condition
\begin{multline}\label{CMarkov}
p(x,t|x_{n-1},t_{n-1};x_{n-2},t_{n-2},\ldots,x_{0},t_{0})\\
=p(x,t|x_{n-1},t_{n-1}),
\end{multline}
where $x,x_{n-1},x_{n-2},\ldots,x_{0}$ are values of a random variable $X$ at times $t>t_{n-1}>t_{n-2}>\ldots>t_0$, respectively. From this equation one obtains that the one-point probabilities change from $t_{n-1}$ to $t$ as
\[
p(x_n,t_n)=\sum_{x_{n-1}}p(x,t|x_{n-1},t_{n-1})p(x_{n-1},t_{n-1}),
\]
and the Chapman-Kolmogorov Equation is satisfied
\begin{multline}\nonumber
p(x,t|x_{n-2},t_{n-2})\\
=\sum_{x_{n-1}}p(x,t|x_{n-1},t_{n-1})p(x_{n-1},t_{n-1}|x_{n-2},t_{n-2}),
\end{multline}
where we have taken $X$ to be a discrete random variable for the sake of simplicity. Therefore the ``transition matrices'' connecting one-point probabilities at a different times are just the conditional probabilities $p(x,t|x_{n-1},t_{n-1})$. Moreover, $p(x,t|x_{n-1},t_{n-1})$ seen as a matrix with indexes $(x,x_{n-1})$ may be non-invertible, however, because its elements are just conditional probabilities, it always satisfies
\begin{align}
& p(x,t|x_{n-1},t_{n-1})\geq0, \label{SM1}\\
& \sum_{x}p(x,t|x_{n-1},t_{n-1})=1. \label{SM2}
\end{align}
This implies that the map defined by the action of this (stochastic) matrix, $p(x',t|x,s)$, with $t>s$, preserves positivity and normalization of any probability distribution, say $p(x)$. Note that this is so even if $p(x)$ is not in the image of the linear map defined by the action of the matrix $p(x,t|x_0,0)$.

Having said that, in the quantum case the role of one-point probabilities and transition matrices is played by density operators $\rho$ and trace preserving and positive maps, respectively (or completely positive maps, depending on whether or not one allows for ancillary degrees of freedom). The key point is that in the classical limit, i.~e. for quantum dynamical situations where coherence does not play any relevant role and density matrices commute one another at any different times, any quantum Markovian definition must fit with the classical Markovian definition. Note that in this case such a quantum dynamics can be completely understood in terms of a completely classical stochastic process. If $V_{t,s}$ is positive (or CP) and trace preserving on ${\rm Im}(\Lambda_s)$ but cannot be extended from ${\rm Im}(\Lambda_s)$ to the whole space keeping these properties, in the classical limit such a dynamics cannot be accepted as Markovian. Indeed, in such a limit, for an appropriate incoherent (diagonal) basis, the evolution is obtained in terms of one-point probabilities given by the diagonal elements of $\rho$ and transition matrices obtained by the matrix elements given by the action of $V_{t,s}$ on the diagonal matrix basis. If $V_{t,s}$ is not extendible, it is clear that the transition matrices of that processes are positive for probability vectors taken from diagonal density matrices in ${\rm Im}(\Lambda_s)$, but not necessarily for any probability vector. Then, the process cannot be accepted as classically Markovian because Eq. \eqref{CMarkov} can be violated! In order words, classical transition matrices obtained from quantum maps in the classical limit should be positive for any probability vector if we intend to understand them as conditional probabilities satisfying Eqs. \eqref{SM1} and \eqref{SM2}. Otherwise, one concludes there is some memory in those processes. Thus, if we want a quantum definition for Markovianity consistent in the classical limit, it has to be possible to extend $V_{t,s}$ to a P (or CP) map onto the whole space.
\item \underline{Memoryless Environment Interpretation}: Another way to motivate the extension of $V_{t,s}$ to $\mathcal{B}(\mathcal{H})$ is the interpretation of a quantum Markovian dynamics as a successive point interaction with a memoryless environment (i.e. the continuous limit of a quantum Markov chain). The idea is that for any interval of time $[0,t]$ we can take $n$ arbitrary intermediate steps $t\geq t_n\geq \ldots \geq t_1 \geq 0$. If the process is CP divisible, the dynamics from $t_{k-1}$ to $t_k$ is given by a CPTP map, and then because the Stinespring representation theorem we may write
\begin{equation}\label{EqCollision}
V_{t_{k},t_{k-1}}(\rho)=\mathrm{Tr}[U(t_{k},t_{k-1})\rho\otimes\omega_E U^\dagger(t_{k},t_{k-1})],
\end{equation}
for some appropriate unitaries $U(t_{k},t_{k-1})$ and state of the environment $\omega_E$. By gauging the dependence on $t_{k}$ and $t_{k-1}$ of $U(t_{k},t_{k-1})$ we can take $\omega_E$ to be the same independently of $t_{k}$. These models are sometimes called collision-like models. Therefore it is clear that the CP-divisibility implies a clear form of the memoryless property: no memory can be kept of the past because it is not possible to distinguish the system evolution from that arising from the interaction with an environment that is reset at each time step.

Note that the fact that a CP-divisible dynamical map admits this kind of collision-like model interpretation does not mean that the underlying joint evolution actually is collision-like. In general, it is possible that correlations between the system and environment are established and kept along the evolution. However, if the system dynamics is CP-divisible, such correlations do not induce to any observable memory effect.

In order to keep this clear interpretation for noninvertible maps, the map $V_{t,s}$ must be extendible from ${\rm Im}(\Lambda_s)$ to the whole space, as Eq. \eqref{EqCollision} requires $V_{t_{k},t_{k-1}}$ to be a CP map not only for $\rho\in{\rm Im}(\Lambda_{t_{k-1}})$, but for the whole space $\mathcal{B}(\mathcal{H})$.
\end{itemize}

\subsection{II.\quad Existence of Hemiticity preserving projectors}
\label{app_A}

Suppose a subspace $M\subset\BH$. Since for any $X\in \BH$ we can write $X=A+iB$ with Hermitian 
\begin{align*}
A=\frac{1}{2}(X+X^\dagger),\quad \text{and}\quad
B=\frac{1}{2i}(X-X^\dagger),
\end{align*}
there exists a set of Hermitian operators spanning $M$. Moreover, this set can be taken to be an orthonormal basis $\{G_\alpha\}$ according to the Hilbert-Schmidt inner product $\tr(G_\alpha^\dagger G_\beta)=\tr(G_\alpha G_\beta)=\delta_{\alpha,\beta}$. This can be achieved, for instance, by applying the Gram-Schmidt orthonormalization procedure (it is easy to check that this transforms a set of Hermitian operators into another set of Hermitian operators). Then a Hermiticity preserving projector onto $M$ is given by
\begin{equation}
\Pi(X)=\sum_{\alpha} \tr(G_\alpha X) G_\alpha.
\end{equation}
\subsection{III.\quad Proof of Lemma 2}
\label{app_B}

Following Jencova \cite{JencovaSM}, for $\dim \mathcal{H}=d$, if the subspace $M\subset\BH$ is generated by positive operators, there exists some positive $\rho\in M$ such that the support of $\rho$ contains the supports of all other elements in $M$. Let us denote by $P$ the projector onto the support of $\rho$, so that $M$ is a subspace of $P\BH P$, and $\rho$ is full-rank in $P\BH P$. Then we define $M':=\rho^{-1/2}M\rho^{-1/2}$ which is an operator system in $P\BH P$. Now, if $\Phi:M\to \BH$ is $d$-positive, $\Phi'(\cdot):=\Phi[\rho^{1/2}(\cdot)\rho^{1/2}]$ is $d$-positive on $M'$, namely, it is CP since $M'$ is an operator system, and the Arveson's extension theorem ensures the existence of a CP extension $\widetilde{\Phi}':\BH\to\BH$. Then, the required CP extension of $\Phi$ is
\[\widetilde{\Phi}=\widetilde{\Phi}'[\rho^{-1/2}(\cdot)\rho^{-1/2}].
\]

\subsection{IV.\quad Proof that $\Pi_{t_1}:=\lim_{\epsilon\rightarrow0^+}V_{t_1,t_1-\epsilon}$ is a  CPTP projector}
\label{app_C}

In order to see that $\Pi_{t_1}$ is well defined, we employ induced norm on the dual space of ${\cal B(H)}$:
\begin{align}
\|V_{t_1,t_1-\epsilon}-&V_{t_1,t_1-\delta}\|=\sup_{Y\neq0}\frac{\|(V_{t_1,t_1-\epsilon}-V_{t_1,t_1-\delta})(Y)\|_1}{\|Y\|_1}\nonumber\\
&=\sup_{X\neq0}\frac{\|(V_{t_1,t_1-\epsilon}-V_{t_1,t_1-\delta})[\Lambda_{t_1-\delta}(X)]\|_1}{\|\Lambda_{t_1-\delta}(X)\|_1}\nonumber\\
&=\sup_{X\neq0}\frac{\|(V_{t_1,t_1-\epsilon}\Lambda_{t_1-\delta}-\Lambda_{t_1})(X)]\|_1}{\|\Lambda_{t_1-\delta}(X)\|_1},
\end{align}
where we have used that, by assumption, $\Lambda_{t_1-\delta}$ is {bijective} for $\delta\in(0,t_1)$. Moreover, since $\|V_{t_1,t_1-\epsilon}-V_{t_1,t_1-\delta}\|=\|V_{t_1,t_1-\delta}-V_{t_1,t_1-\epsilon}\|$ we can take $\delta\geq \epsilon$ without loss of generality. Then by writing $\Lambda_{t_1}=V_{t_1,t_1-\epsilon}\Lambda_{t_1-\epsilon}$ and using that $V_{t_1,t_1-\epsilon}$ is a TP contraction for $\epsilon\in(0,t_1)$ we arrive at
\begin{widetext}
\begin{align}
\|V_{t_1,t_1-\epsilon}-V_{t_1,t_1-\delta}\|&=\sup_{X\neq0}\frac{\|(V_{t_1,t_1-\epsilon}\Lambda_{t_1-\delta}-V_{t_1,t_1-\epsilon}\Lambda_{t_1-\epsilon})(X)]\|_1}{\|\Lambda_{t_1-\delta}(X)\|_1}\nonumber\\
&\leq\|V_{t_1,t_1-\epsilon}\|\sup_{X\neq0}\frac{\|(\Lambda_{t_1-\delta}-\Lambda_{t_1-\epsilon})(X)]\|_1}{\|\Lambda_{t_1-\delta}(X)\|_1}\nonumber\\
&\leq\sup_{X\neq0}\frac{\|(\Lambda_{t_1-\delta}-\Lambda_{t_1-\epsilon})(X)]\|_1}{\|\Lambda_{t_1-\delta}(X)\|_1}.
\end{align}
Since $\lim_{\epsilon\rightarrow0^+}\Lambda_{t_1-\epsilon}=\Lambda_{t_1}$ (specifically, the map $t\mapsto \Lambda_t$ is norm continuous) the family $V_{t_1,t_1-\epsilon}$ is Cauchy convergent as $\epsilon\rightarrow0^+$. Therefore the limit $\lim_{\epsilon\rightarrow0^+}V_{t_1,t_1-\epsilon}$ is convergent as the dual space of ${\cal B(H)}$ is a Banach space with the induced norm.

Similar manipulations show that $\Pi_{t_1}$ is idempotent:
\begin{align}
\|\Pi_{t_1}^2-\Pi_{t_1}\|&=\sup_{Y\neq0}\lim_{\epsilon,\delta\rightarrow0^+}\frac{\|(V_{t_1,t_1-\epsilon}V_{t_1,t_1-\delta}-V_{t_1,t_1-\delta})(Y)\|_1}{\|Y\|_1}\nonumber\\
&=\sup_{X\neq0}\lim_{\epsilon,\delta\rightarrow0^+}\frac{\|(V_{t_1,t_1-\epsilon}\Lambda_{t_1}-\Lambda_{t_1})(X)\|_1}{\|\Lambda_{t_1-\delta}(X)\|_1}\nonumber\\
&\leq\sup_{X\neq0}\lim_{\epsilon,\delta\rightarrow0^+}\|V_{t_1,t_1-\epsilon}\|\frac{\|(\Lambda_{t_1}-\Lambda_{t_1-\epsilon})(X)\|_1}{\|\Lambda_{t_1-\delta}(X)\|_1}\leq\sup_{X\neq0}\lim_{\epsilon,\delta\rightarrow0^+}\frac{\|(\Lambda_{t_1}-\Lambda_{t_1-\epsilon})(X)\|_1}{\|\Lambda_{t_1-\delta}(X)\|_1}=0.
\end{align}
\end{widetext}
On the other hand, from $\Lambda_{t_1}=V_{t_1,t_1-\epsilon}\Lambda_{t_1-\epsilon}$ and again due to the fact that $\Lambda_{t_1-\epsilon}$ is {bijective} for $\epsilon\in(0,t_1)$, we conclude that the range of $V_{t_1,t_1-\epsilon}$ is the same as the range of $\Lambda_{t_1}$, and so the image of $\Pi_{t_1}$ is ${\rm Im}(\Lambda_{t_1})$. Moreover, since $V_{t_1,t_1-\epsilon}$ is positive and TP for $\epsilon\in(0,t_1)$, we conclude that $\Pi_{t_1}$ is a TP and positive projection onto ${\rm Im}(\Lambda_{t_1})$.\par
\hfill $\Box$

\subsection{V.\quad Example: Relaxation to the equilibrium state}
\label{app_D}

Consider the dynamical map
\begin{equation}\label{}
  \Lambda_t \rho = [1- F(t)] \rho + F(t) \omega {\rm Tr}(\rho) , \quad t\geq0
\end{equation}
where $\omega$ is an equilibrium density operator satisfying $\Lambda_t \omega = \omega$. $\Lambda_0=\mathds{1}$ implies $F(0)=0$ and if $0 \leq F(t) \leq 1$ the map is CPTP. In addition, it is invertible iff $F(t)<1$.
Suppose that for some time $t_*$ one has $F(t_*)=1$, i.~e. $\Lambda_{t_*}\rho  = \omega {\rm Tr}(\rho)$ is not invertible.
The dynamical map $\{\Lambda_t\}_{t\geq0}$ is divisible iff $   \Lambda_{t} = \Lambda_{t_*}$
for $t \geq t_*$, or, equivalently, iff $F(t)=1$ for $t \geq t_*$. One has for the propagator $V_{t,s} = \Lambda_t \Lambda_s^{-1}$ for $s < t_*$, with
$ \Lambda_s^{-1}\rho=\frac{\rho}{1-F(s)}-\frac{F(s)}{1-F(s)}\omega {\rm Tr}(\rho)$.
The projector $\Pi_{t_*}$ is given by
\begin{equation*}\label{}
\Pi_{t_*}\rho=\lim_{\epsilon\rightarrow0^+}V_{t_{*},t_{*}-\epsilon}\rho = \lim_{\epsilon\rightarrow0^+}\Lambda_{t_*}\Lambda_{t_{*}-\epsilon}^{-1}\rho=\omega{\rm Tr}(\rho) .
\end{equation*}
So we can take $V_{t,s} = \Pi_{t_*}$ for $s \geq  t_*$. Again, divisibility means that if $\Lambda_t\rho$ relaxes to $\omega$ at finite time $t_*$ it stays at $\omega$ forever. CP-divisibility is equivalent to divisibility plus the monotonicity condition $\dot{F}(t) \geq 0$ for $t< t_*$.
Finally, one finds for the generator
\begin{equation}\label{}
  \mathcal{L}_t\rho= \gamma(t) (\omega {\rm Tr}(\rho) - \rho) ,
\end{equation}
with $\gamma(t) = \frac{\dot{F}(t)}{1-F(t)} \geq 0$ for $t < t_*$. This formula does not work for $t \geq t_*$ due to $F(t)=1$. Hence, for $t \geq t_*$ the rate $\gamma(t)$ is arbitrary (provided that $\int_0^t \gamma(\tau)d\tau = \infty$ for $t \geq t_*$).

\subsection{VI.\quad Proof of Theorem 5}
\label{app_E}
We first note that if 
\begin{equation}\label{dCPdiv2SM}
\frac{d}{dt}\|(\mathds{1}_{d+1} \otimes\Lambda_t)({\rho}_1-{\rho}_2)\|_1\leq0
\end{equation}
is satisfied, then $\{\Lambda_t\}_{t\geq0}$ is divisible. Indeed, otherwise there exists ${X}'\in\mathcal{B}(\mathcal{H}' \ot \mathcal{H})$ such that $(\mathds{1}_{d+1}\otimes \Lambda_s)({X}')=0$ and $(\mathds{1}_{d+1}\otimes  \Lambda_t)({X}') \neq 0$ for some pair $t>s$. Here $\mathcal{H}'$ is the $d+1$ dimensional Hilbert space with orthonormal basis $\{|i\rangle\}_{i=1}^{d+1}$ where $\{|i\rangle\}_{i=1}^{d}$ is a basis for $\mathcal{H}$.

Since $\mathds{1}_{d+1}\otimes \Lambda_s$ is TP, we get ${\rm Tr}({X}')=0$, so that ${X}'$ can be written (up to some multiplicative constant) as a difference of two density operators ${X}'={\rho}_1-{\rho}_2$. Therefore, for a non-divisible map we have
\begin{equation}
\|(\mathds{1}_{d+1}\otimes  \Lambda_t) ({\rho}_1-{\rho}_2)\|_1 > 0=\|(\mathds{1}_{d+1}\otimes \Lambda_s)({\rho}_1-{\rho}_2)\|_1,
\end{equation}
and hence the inequality \eqref{dCPdiv2SM} is violated, proving our assertion that $\{\Lambda_t\}_{t\geq0}$ is divisible, hence $\Lambda_t=V_{t,s}\Lambda_s$. 

Thus, by \eqref{dCPdiv2SM} we get for ${\rm Tr}(X')=0$,
\begin{equation}
\frac{d}{dt}\|(\mathds{1}_{d+1} \otimes\Lambda_t)(X')\|_1\leq0.
\end{equation}
Hence $\mathds{1}_{d+1}\otimes V_{t,s}$ is a TP contraction on the subspace of traceless operators in ${\rm Im}(\mathds{1}_{d+1}\otimes\Lambda_{s})$. 

We next use this to show $\mathds{1}\otimes V_{t,s}$ is a TP contraction on all ${\rm Im}(\mathds{1}\otimes\Lambda_{s})$ \cite{footnote2}.

Let ${Y}\in{\rm Im}(\mathds{1}\otimes\Lambda_{s})$, ${Y}=(\mathds{1}\otimes\Lambda_{s})({X})$ with ${X}\in\mathcal{B}(\mathcal{H}\otimes \mathcal{H})$. Then, similarly to \cite{BOGNASM}, consider
\begin{equation}
\Delta:={X}-{\rm Tr}({X})|d+1\rangle \langle d+1|\otimes \rho_S\in\mathcal{B}(\mathcal{H}'\otimes\mathcal{H}),
\end{equation}
where $\mathcal{H}'$ is as above, and $\rho_S$ a density operator. Then $\Delta$ is traceless, hence so is $(\mathds{1}_{d+1}\otimes \Lambda_{s})(\Delta)$. Since $\mathds{1}_{d+1}\otimes V_{t,s}$ is a TP-contraction on the subspace of traceless operators in ${\rm Im}(\mathds{1}_{d+1}\otimes\Lambda_{s})$, we get
\begin{equation}\label{dCPdiv2P}
\|(\mathds{1}_{d+1}\otimes V_{t,s}\Lambda_{s}) (\Delta)\|_1\leq \|(\mathds{1}_{d+1}\otimes\Lambda_{s})(\Delta)\|_1.
\end{equation}

Now ${X}\in \mathcal{B}(\mathcal{H}\otimes\mathcal{H})$, and $|d+1\rangle$ is orthogonal to $\mathcal{H}$, so $X$ and $|d+1\rangle \langle d+1|\otimes \rho_S$ have orthogonal supports. Therefore $Y=(\mathds{1}_{d+1}\otimes\Lambda_s)(X)$ and $(\mathds{1}_{d+1}\otimes\Lambda_s)(|d+1\rangle \langle d+1|\otimes \rho_S)$ have orthogonal supports too. Since $\Lambda_{t}$ is a CPTP map, $\Lambda_t(\rho_S)$ is also a density operator. Hence we get
\begin{align*}
\|(\mathds{1}_{d+1}\otimes V_{t,s}\Lambda_{s})(\Delta)\|_1&=\|(\mathds{1}\otimes V_{t,s})({Y})\|_1\\
&\ +\|{\rm Tr}({X})|d+1\rangle \langle d+1|\otimes \Lambda_t(\rho_S)\|_1\\
&=\|(\mathds{1}\otimes V_{t,s})({Y})\|_1+|{\rm Tr}({X})|
\end{align*}
and
\[
\|(\mathds{1}_{d+1}\otimes\Lambda_{s})(\Delta)\|_1=\|{Y}\|_1+|{\rm Tr}({X})|.
\]
It follows from \eqref{dCPdiv2P} that
\begin{align*}
\|(\mathds{1}_{d+1}\otimes V_{t,s})(Y)\|_1&=\|(\mathds{1}\otimes V_{t,s}\Lambda_s)(\Delta)\|_1-|{\rm Tr}({X})|\\
&\leq \|(\mathds{1}\otimes \Lambda_s)(\Delta)\|_1-|{\rm Tr}({X})|\\
&=\|Y\|_1,
\end{align*}
which proves that $\mathds{1}_{d+1}\otimes V_{t,s}$ is a TP-contraction on ${\rm Im}(\mathds{1}\otimes\Lambda_s)$. Therefore, the theorem follows from Theorems 3 and 4.\par
\hfill $\Box$
\vspace{-0.5cm}


\begin{thebibliography}{1}
\bibitem[S3]{footnote2} Note that throughout the text we have omitted the subindex $d=\dim(\mathcal{H})$ for the identity on $\BH$ (where $\Lambda_t$ acts on), $\mathds{1}\equiv\mathds{1}_d$.
\end{thebibliography}
\end{document}